\documentclass[conference]{IEEEtran}
\usepackage[utf8x]{inputenc}

\usepackage{amsmath}
\usepackage{amsfonts}
\usepackage{amssymb}

\usepackage{subfigure}
\usepackage{tikz}
\usetikzlibrary{calc,3d}
\usepackage{listings}

\usepackage[shell]{gnuplottex}

\title{Avoiding Serialization Effects in Data / Dependency aware Task Parallel Algorithms for
Spatial Decomposition. }

\author{\IEEEauthorblockN{Christoph Niethammer \\ and Colin W. Glass \\ and José Gracia}
\IEEEauthorblockA{High performance computing center Stuttgart\\
University of Stuttgart\\
Email: \{niethammer,glass,gracia\}@hlrs.de}
}

\begin{document}

\maketitle


\begin{abstract}
Spatial decomposition is a popular basis for parallelising code. Cast in the frame
of task parallelism, calculations on a spatial domain can be treated as a task. If
neighbouring domains interact and share results,  access to the specific data needs
to be synchronized to avoid race conditions. This is the case for a variety of applications,
like most molecular dynamics and many computational fluid dynamics
codes. Here we present an unexpected problem which can occur in dependency-driven
task parallelization models like StarSs: the tasks accessing a specific spatial domain are
treated as interdependent, as dependencies are detected automatically via memory
addresses. Thus, the order in which tasks are generated will have a severe impact on
the dependency tree. In the worst case, a complete serialization is reached and no two
tasks can be calculated in parallel. We present the problem in detail based on an
example from molecular dynamics, and introduce a theoretical framework to calculate the degree of
serialization. Furthermore, we present strategies to avoid this unnecessary problem.
We recommend treating these strategies as best practice when
using dependency-driven task parallel programming models like StarSs on such scenarios.
\end{abstract}

\section{Introduction}
\label{sec:introduction}

Parallel programming is the key to modern day computing. A variety of new
parallel programming models have come up in recent years
, facilitating the
often difficult task of parallelizing code. One of them is StarSs, a dependency-aware task
based programming model which automatically extracts program parallelisation at
the task level \cite{Planas2009}. This is achieved by comparing memory addresses
of input and output variables -- thereby detecting data dependencies between tasks.
Thus, the programmer need not deal with the actual parallelisation, but simply specifies
chunks of work to be treated as tasks. The StarSs programming model is described in section~\ref{seq:starss-programming-model}

In this paper,  we present a scenario from everyday HPC computing -- spatial
decomposition with result sharing between neighbouring areas -- where this
elegant approach is led astray, resulting in a catastrophic failure
of the intended parallelisation.
Section~\ref{sec:scope-of-problem} outlines the scope of applications, where this problem can occur, while section \ref{sec:example} describes the actual case studied.
 In section \ref{sec:straight-forward-implementation} a theoretical framework on the speedup of a basic implementation as found in literature
is introduced and section \ref{sec:improved_strategies} presents strategies to achieve the best possible parallelization by adhering to
simple rules regarding the chronological ordering of task creation.
The theoretical speedup and the improved strategies are both verified experimentally in section \ref{sec:evaluation}.

\section{Scope of the addressed problem}
\label{sec:scope-of-problem}

Many codes make use of spatial decomposition to achieve parallelism. In some cases,
data of neighbouring areas is on a read-only basis. Since no data is written to neighbouring
areas, all areas are independent from each other and therefore the risk of unwarranted dependencies
not given. However, there are many cases where results are partly equivalent for
neighbouring areas. To avoid calculating the same values twice, they are written
to both areas. Examples herefor are e.g. Molecular Dynamics (MD) (Newton's third law) and finite volume methods
(flux across boundaries). These are examples of cases relevant to this paper. In both the abovementioned
cases, i.e. force on a molecule and flux to/from a cell, only the net total is relevant. Therefore, the
contributions of different neighbours are summed up, hereby avoiding unnecessary memory
consumption. Thus, different neighbours will write to the same memory. Race conditions
need to be avoided by preventing simultaneous write access.
If we assume we treat the calculation of one area as one task in StarSs, different neighbours
wanting to write to the same memory address will lead to the automatic detection of dependencies
between these neighbours.
Obviously, it is sufficient to avoid simultaneous write access. However, StarSs does not know that. Therefore,
not only all neighbours wanting to write to one area need to wait, but all their
neighbours have to wait on them in turn and so on. This leads to unnecessary serialization. The degree of
serialization depends on the exact nature of the areas, definition of tasks and most important, on the order of task
generation. As we will show on the case of the link-cell algorithm for MD, the standard 
implementation from literature leads in the worst case to complete serialization, or in other words, a maximum speedup of 1.
By adhering to simple strategies, the maximum speedup can  be increased significantly.


\section{The StarSs programming model}
\label{seq:starss-programming-model}
StarSs is a dependency aware task based parallel programming model. The basic idea is the extraction of parallelism out of a serial program  using input and output dependencies.  This frees the programmer from the difficult task of identifying parallel parts in his program, leaving it to the StarSs runtime.  Currently StarSs implementations are available for C/C++ (Ompss) and Fortran (SMPSs).

In StarSs tasks are generated from functions or methods.  To declare a function or method as a task, code annotations in form of pragmas (C/C++) or comments (Fortran) similar to OpenMP \cite{OpenMP} are used.  For each task the input and output parameters have to be defined.  In C/C++ this requires additional pragma parameters, while in Fortran the parameters' intent clauses are used.  During runtime the actual dependencies between tasks are determined by  the memory addresses of the passed parameters.   In figure~\ref{code:starss-example} an example C~program is shown.

\begin{figure}[ht]
\begin{lstlisting}
#pragma css task input(N,a,b) output(c)
void vec_add(int N,
        double *a, double *b, double *c)

#pragma css task input(N,a,b) output(s)
void scalar_product(int N,
        double *a, double *b, double *s)

#pragma css task input(N,j) output(a)
void unit_vec(int N, int j, double *a)

int main() {
int N;
double s, *a, *b, *c, *d;
...
#pragma css start
    unit_vec(N, 0, a);
    unit_vec(N, 1, b);
    unit_vec(N, 2, c);
    vec_add(N, c, b, c);
    vec_add(N, a, b, d);
    scalar_product(N, c, d, &s);
#pragma css finish
...
}
\end{lstlisting}
\caption{Simple C program calculating $s = \langle\vec{e_2}+\vec{e_1} , \vec{e_o}+\vec{e_1}\rangle$ for the N dimensional unit vectors $e_i$ with StarSs.}
\label{code:starss-example}
\end{figure}

The StarSs runtime starts the program in a master thread.  Every call to a function or method with a task annotation results in the creation of a task with the respective dependencies. This task is  added to the task execution list and the master thread continues. Worker threads created by the runtime look through this list and search for tasks which have no unmet input dependencies. These tasks  are ready to be executed.

\begin{figure}[ht]
\centering
\begin{tikzpicture}[scale=1]
\tikzstyle{every node}=[fill=yellow!60,circle,inner sep=1pt]
\node at ( 0, 2) [fill=green!40] (n1){1};
\node at ( 1, 2) [fill=green!40] (n2) {2};
\node at ( 2, 2) [fill=green!40] (n3) {3};
\node at ( 0.5, 1) [fill=yellow!40] (n4){4};
\node at ( 1.5, 1) [fill=yellow!40] (n5) {5};
\node at ( 1, 0) [fill=red!40] (n6) {6};
\draw[->] (n1.south) -- (n4.north);
\draw[->] (n2.south) -- (n4.north);
\draw[->] (n2.south) -- (n5.north);
\draw[->] (n3.south) -- (n5.north);
\draw[->] (n4.south) -- (n6.north);
\draw[->] (n5.south) -- (n6.north);
\end{tikzpicture}
\caption{Task graph for the example program given in \ref{code:starss-example}. Vertices represent tasks where the color stands for the task function. Edges represent dependencies. The direction of an edge is from output to input parameters and is always oriented downwards in the DAG. Numbers inside the vertices denote the chronological order of task creation in the program.}
\label{fig:starss-example-dag}
\end{figure}
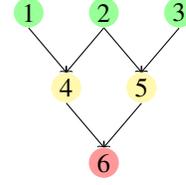

The dependencies between the tasks can be represented by a directed acyclic graph (DAG) which is shown in figure~\ref{fig:starss-example-dag}.  Tasks are the vertices of the DAG and edges are the dependencies. 

The StarSs model includes additional statements for runtime initialization/finalization and for the synchronization between the master and worker threads.  For a detailed description we refer to the SMPSs manual\cite{SMPSs-2.4}.

\section{Example: Molecular Dynamics}
\label{sec:example}

A standard method to speed up Molecular Dynamics calculations for short range potentials is the link-cell~(LC)
method~\cite{Griebel1998}. It allows to reduce the complexity of the search for interaction partners from $\mathcal{O}(N^2)$ to $\mathcal{O}(N)$ for $N$ particles\footnote{A particle is small localized object with properties, e.g. an atom or a molecule.}.
The basic idea behind the LC method is to decompose the simulation area into smaller cells, which are of the size of the cutoff radius ($r_\text{c}$).
Therefore, while searching for the interaction partners of one particle, only particles within the same and the neighbouring
cells need to be considered. All other particles are beyond the cutoff radius~$r_\text{c}$.

\begin{figure}[h]
  \centering
  \subfigure{ \begin{tikzpicture}[scale=0.75]
  \draw[fill=red,opacity=.8]  (1,2) rectangle(2,3);
  \draw[opacity=.3,fill=orange]  (1,1) rectangle(2,2);
  \draw[opacity=.3,fill=orange]  (2,1) rectangle(3,4);
  \draw[fill=black] (1.8,2.2) circle (0.1);
  \draw (1.8,2.2) circle (1);
  \draw[<->,color=black] (1.8,2.2) --  node[anchor=north]{$r_\text{c}$}+(30:1);
  \draw grid (5,5);
  \draw [<->] (1,-.25) to node[below] {$r_\text{c}$} +(1,0);
\end{tikzpicture} \label{fig:2D-stencil}} \hspace{8pt} \subfigure[][]{ \begin{tikzpicture}[x  = {(-0.5cm,-0.5cm)},
                    y  = {(0.9659cm,-0.25882cm)},
                    z  = {(0cm,1cm)},
                    scale = 0.75,
                    color={lightgray}]
  \tikzset{facestyle/.style={fill=lightgray,draw=black,very thin,line join=round}}
  \tikzset{outer/.style={opacity=.8,draw=black,very thin,line join=round}}
  \tikzset{outer2/.style={opacity=.3,fill=orange,draw=black,very thin,line join=round}}
  \tikzset{inner/.style={fill=red,opacity=.8,draw=black,very thin,line join=round}}

  \newcommand{\drawCell}[4] {
      \begin{scope}[canvas is zy plane at x={#1+1}]
        \path[#4] (#3,#2) rectangle +(1,1);
      \end{scope}
      \begin{scope}[canvas is zx plane at y={#2+1}]
        \path[#4] (#3,#1) rectangle +(1,1);
      \end{scope}
      \begin{scope}[canvas is yx plane at z={#3+1}]
        \path[#4] (#2,#1) rectangle +(1,1);
      \end{scope}
  }

  \begin{scope}[canvas is zx plane at y=0]
    \path[outer] (1,0) -- +(0,3);
  \end{scope}

  \drawCell{0}{0}{1}{outer}
  \drawCell{0}{0}{2}{outer}
  \drawCell{0}{0}{3}{outer}
  \drawCell{0}{1}{1}{outer}
  \drawCell{0}{1}{2}{outer}
  \drawCell{0}{1}{3}{outer}
  \drawCell{0}{2}{1}{outer2}
  \drawCell{0}{2}{2}{outer2}
  \drawCell{0}{2}{3}{outer2}

  \drawCell{1}{0}{1}{outer}
  \drawCell{1}{0}{2}{outer}
  \drawCell{1}{0}{3}{outer}
  \drawCell{1}{1}{1}{outer2}
  \drawCell{1}{1}{2}{inner}
  \drawCell{1}{1}{3}{outer}
  \drawCell{1}{2}{1}{outer2}
  \drawCell{1}{2}{2}{outer2}
  \drawCell{1}{2}{3}{outer2}

  \drawCell{2}{0}{1}{outer}
  \drawCell{2}{0}{2}{outer}
  \drawCell{2}{0}{3}{outer}
  \drawCell{2}{1}{1}{outer2}
  \drawCell{2}{1}{2}{outer2}
  \drawCell{2}{1}{3}{outer2}
  \drawCell{2}{2}{1}{outer2}
  \drawCell{2}{2}{2}{outer2}
  \drawCell{2}{2}{3}{outer2}
\end{tikzpicture} \label{fig:3D-stencil}}
  \caption{Stencils for 2D~\subref{fig:2D-stencil} and 3D~\subref{fig:3D-stencil} as they are used in the link-cell algorithm with actio=reactio. The cell dimensions are equal to the cutoff radius $r_\text{c}$ of the interaction potential. The red cell is the center of the stencil, the orange cells are the neighbour cells. Red and orange cells are both updated.}
  \label{fig:neighbor-stencils}
\end{figure}
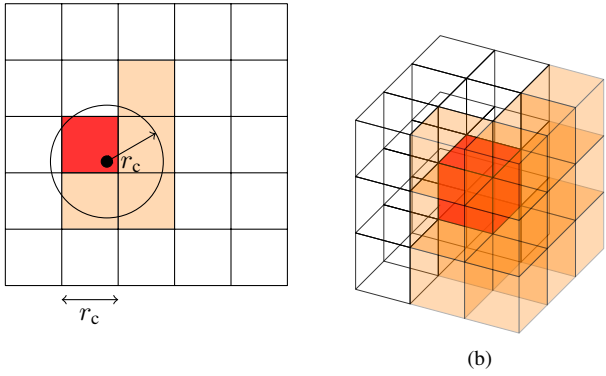

The number of required computations can be reduced further by considering Newton's third law (action = reactio). For a given particle, only half the interactions need to be calculated. When the interaction of a particle A with another particle B is calculated, the result of the respective interaction of particle B with particle A is already known, so the second calculation can be omitted. Clearly, this requires writing to the own memory and to the memory of the interaction partner at the time of force calculation. But by this only half of the neighbouring cells have to be considered during calculation.
The  stencils\footnote{A stencil is a geometric pattern, relative to the point of interest.} for the 2D and 3D case are shown in Fig.~\ref{fig:neighbor-stencils}.

In the following we consider the algorithm on the cell level. We define the term cell-cell interaction as the calculation of all interactions of the particles in one cell with all particles in another cell. A cell interacting with itself also qualifies as a cell-cell interaction. Where we need to distinguish explicitly between the interaction within one cell and between two differing cells, we denote this by the terms intra- and inter-cell interaction respectively.

Although in principal it does not make any difference, in which order the interactions are
calculated, the LC algorithm as described in literature follows a given procedure\cite{Griebel1998}:
it starts with a cell and calculates the 
intra-cell interaction, followed by the inter-cell interactions, before moving on to the
next cell.

Bringing this algorithm to the StarSs programming model is straight forward:
every cell-cell interaction is a task. This provides sufficient work load for each task,
while delivering a lot of tasks compared to
treating all cell-cell interactions of one cell as a single task.

As we update both cells involved in a task, they are specified as inout dependency. This will protect the
particle data from being updated by multiple tasks/threads simultaneously.
The StarSs runtime will then create a directed acyclic graph (DAG) of the specified tasks taking into account the given dependencies and execute the program with the extracted parallelism.

As we will show in the following sections, the degree
of parallelism will depend on the order of task creation and therefore on the way
the stencil is evaluated for all elements.

\section{Examination of the straightforward implementation}
\label{sec:straight-forward-implementation}

In the StarSs task DAG the critical path length is the longest path from the
start to the end of a program. The critical path length~$t_\text{cp}$
determines the maximal possible speedup~$S$ of a program, as the tasks in the path have to
be executed in sequential order:
\begin{equation}
  S = \frac{t_\text{s}}{t_\text{cp}}
  \label{eq:speedup}
\end{equation}
where $t_\text{s}$ is the sequential execution time of the program.

For the following considerations we assume that all tasks have equal
execution time. Under this assumption, the critical path length can be given by
the number of executed tasks instead of the overall duration of their
execution. Looking at the DAG, we further neglect some of the dependencies as the static
nature and periodicity of the stencil allows it.

\begin{figure}[h]
  \centering
  \includegraphics[width=0.5\linewidth]{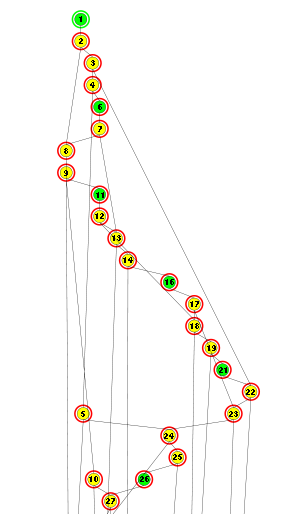}
  \hspace{8pt}
  \begin{tikzpicture}[scale=0.5]
  \tikzstyle{every node}=[fill=yellow!60,circle,inner sep=1pt]
  \node[fill=green!40]{1}
  child {node{2}
  child {node{3}
  child {node{4}
  child {node{5} }
  child {node[fill=green!40]{6}
    child {node{7}
      child {node{8}
        child {node{9}
          child {node{10} }
          child {node[fill=green!40]{11}
  } } } } } } } };
  \end{tikzpicture}

  \caption{DAG of the na\"{\i}ve implementation~(see Fig~\ref{fig:2D-stencil-naiive}) for the 2D-stencil extracted during program execution with Temanejo and the simplified DAG as we use it for the theoretical prediction of the critical path length (red fringe). Intra-cell interaction tasks are depicted in green,  these for inter-cell in yellow.}
  \label{fig:2D-MD-stencil-DAG}
\end{figure}
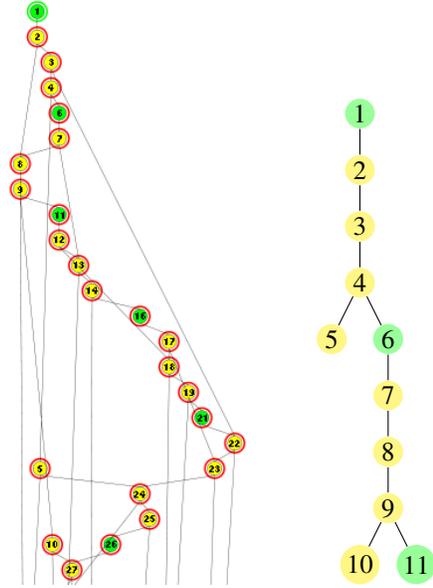

As discussed earlier, the link-cell as described in literature calculates the intra-cell
and then the inter-cell interactions.
The stencil is moved over the cells first
in x, then y and last in z direction.
We will refer to this as the basic implementation.
As we have
four inter-cell interactions there exist in total $4! = 24$ different 
permutations of the execution order of the tasks.

The achievable speedup can be expressed using three numbers: The number of stencil executions for the whole domain $k$, the number of tasks in the single stencil $n$ and the stencil displacement, $\Delta$, which is the number of interactions to be calculated before the interaction with the next neighbour can be done:
\begin{equation}
  S(\Delta,k,n) = \frac{n k}{n + \Delta(k - 1)} \; .
  \label{eq:speedup-stencil}
\end{equation}
In the limit $k \rightarrow \infty$ we get a speedup of
\begin{equation}
S_\text{max}(\Delta,n) = \lim_{k\rightarrow \infty} S(\Delta,k,n) = \frac{n}{\Delta} \; .
  \label{eq:speedup-stencil-max}
\end{equation}
For the basic implementation, $\Delta$ for the 2D stencil varies (depending on the permutation) between $2$ and $5$. It is $2$ if the neighbour in x~direction follows the intra-cell evaluation and $5$ if it is the last interaction evaluated. For the 3D case $\Delta$ is within $[2,14]$ respectively.

Therefore, in both cases the worst case is thus $\Delta = n$ and so by \eqref{eq:speedup-stencil-max} no speedup is
achieved at all! The best case is \mbox{$S_\text{max}(2,5) = 2.5$} for 2D and \mbox{$S_\text{max}(2,14) = 7$} for 3D.

\begin{figure}[h]
  \centering
\subfigure[][]{
  \begin{tikzpicture}
  \draw[fill=black!50]  (0,0) rectangle+(1,1);
  \draw[fill=black!75]  (0,1) rectangle+(1,1);
  \draw[fill=black!50]  (1,0) rectangle+(1,1);
  \draw[fill=black!50]  (1,1) rectangle+(1,1);
  \draw[fill=black!50]  (1,2) rectangle+(1,1);
  \draw (0.5,1.5) node{1} ++(0,-1) node {2} ++(1,0) node{3} ++(0,1) node{4}
++(0,1) node{5};
  \end{tikzpicture}
  \label{fig:2D-stencil-naiive}
}
  \subfigure[][]{
  \begin{tikzpicture}
  \draw[fill=black!50]  (0,0) rectangle+(1,1);
  \draw[fill=black!75]  (0,1) rectangle+(1,1);
  \draw[fill=black!50]  (1,0) rectangle+(1,1);
  \draw[fill=black!50]  (1,1) rectangle+(1,1);
  \draw[fill=black!50]  (1,2) rectangle+(1,1);
  \draw (0.5,1.5) node{1} ++(0,-1) node {2} ++(1,0) node{3} ++(0,1) node{5}
++(0,1) node{4};
  \end{tikzpicture}
  \label{fig:2D-stencil-bad}
}
\subfigure[][]{
  \begin{tikzpicture}
  \draw[fill=black!50]  (0,0) rectangle+(1,1);
  \draw[fill=black!75]  (0,1) rectangle+(1,1);
  \draw[fill=black!50]  (1,0) rectangle+(1,1);
  \draw[fill=black!50]  (1,1) rectangle+(1,1);
  \draw[fill=black!50]  (1,2) rectangle+(1,1);
  \draw (0.5,1.5) node{5} ++(0,-1) node {4} ++(1,0) node{2} ++(0,1) node{1}
++(0,1) node{3};
  \end{tikzpicture}
  \label{fig:2D-stencil-opt}
}
  \caption{Different possible execution orders for cell interactions in the
2D-Stencil
  and their theoretical speedups:
    \subref{fig:2D-stencil-naiive} na\"{\i}ve: \mbox{$S_\text{max} =1.25$},
    \subref{fig:2D-stencil-bad} bad/worst case: \mbox{$S_\text{max} =1$},
    \subref{fig:2D-stencil-opt} optimal stencil: \mbox{$S_\text{max} = 5$}}
  \label{fig:2D-stencil-dag-speedups}
\end{figure}
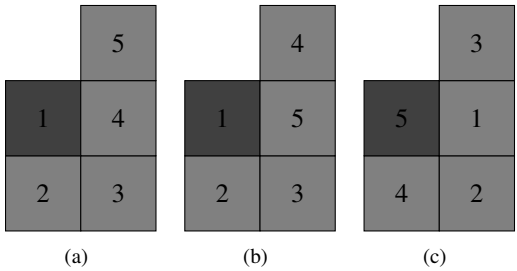

Figure~\ref{fig:2D-MD-stencil-DAG} shows a part of the real and
simplified DAG for a suboptimal basic 2D stencil, which we refer to as the "na\"{\i}ve"\footnote{Na\"{\i}ve because it is the geometrically most appealing stencil, but has a pretty terrible speedup.} version ~(see Fig~\ref{fig:2D-stencil-naiive}), applied
to a $5 \times 5$ cell domain
with periodic boundary conditions. The real DAG was extracted with
Temanejo\cite{Brinkmann2011}.

However, if we disregard the basic stencil implementation and evaluate the cell-cell interaction with the neighbour in x~direction first (see Fig.~\ref{fig:2D-stencil-opt}), then $\Delta = 1$ and we get a DAG which can be seen as a
stencil pipeline and \eqref{eq:speedup-stencil} becomes the formula for
the speedup of a pipeline
\begin{equation}
  S = \frac{n k}{n + k -1}
  \label{eq:speedup-pipeline}
\end{equation}
where now $n$ is the number of operations and $k$ is the pipeline depth. Here
$k$ corresponds to the number of stencils which will be executed concurrently until the
first stencil finishes. This is the optimal version for the stencil implementation and
will at most result in a speedup of $k$. In 2D $k$ equals $5$, in 3D $14$, therefore even the optimal stencil implementation cannot be parallelized across an entire many-core node with $16$~cores or more.

\section{Improved strategies}
\label{sec:improved_strategies}
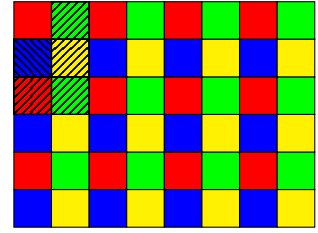
\begin{figure}
  \centering
  \usetikzlibrary{patterns}
\begin{tikzpicture}[scale=0.5]
\foreach \x in {0,2,...,6}{
	\foreach \y in {0,2,...,4} {
		\draw[fill=red] (\x,\y)++(0,1.0) rectangle +(1.0,1.0);
		\draw[fill=blue] (\x,\y)++(0,0.) rectangle +(1.0,1.0);
		\draw[fill=green] (\x,\y)++(1.0,1.0) rectangle +(1.0,1.0);
		\draw[fill=yellow] (\x,\y)++(1.0,0) rectangle +(1.0,1.0);
	}
	\draw[pattern=north west lines] (0,4) rectangle +(1.0,1.0);
	\draw[pattern=north east lines] (0,4)++(1.0,1.0) rectangle +(1.0,1.0);
	\draw[pattern=north east lines] (0,4)++(1.0,0) rectangle +(1.0,1.0);
	\draw[pattern=north east lines] (0,4)++(1.0,-1.0) rectangle +(1.0,1.0);
	\draw[pattern=north east lines] (0,4)++(0,-1.0) rectangle +(1.0,1.0);
}
\end{tikzpicture}
  \caption{2D coloring scheme: Executing the stencil on the cells of one color does not introduce dependencies between tasks.}
  \label{fig:coloring-scheme-2D}
\end{figure}

As we have seen, the stencil implementation, with an outer loop over the cells and an inner loop over the stencil, runs into often severe speedup limitations, due to the unnecessarily generated dependencies. To overcome these limitations, different strategies can be applied:

\begin{enumerate}
  \item The first way to improve the scalability is parallelizing the stencil execution by changing loop ordering. We will refer to this as ``loop exchanged'' (loopex) version later on.
Instead of using an outer loop over all cells and an inner loop over the stencil pattern, the loops are exchanged and therefore the outer loop is over the stencil pattern and the  inner loop over the cells. This strategy will reduce the dependencies between the tasks dramatically and allows the parallel execution of at least one row at once.

A smaller pitfall of this method is the direct dependencies in the loop directions (x,y and z) which can introduce again a serialization when it comes to neighbour dependencies. To circumvent this, a coloring scheme is used: Two-strided loops go over the domain preventing locks due to dependencies in the forward cell interaction steps for each of the directions x, y and z.  For the coloring scheme used in the 2D example see Fig.~\ref{fig:coloring-scheme-2D}.  Note that a simple coloring scheme without loop reordering cannot extract the same amount of parallelism.  This can be trivially seen by looking at figure~\ref{fig:coloring-scheme-2D}.

The implementation of this optimization strategy clearly requires a bit of code overhead compared to the original implementation and especially the coloring scheme is an error prone part.

 \item The second approach is a stencil with task nesting (currently only supported by
Ompss \cite{Duran2011}). In this version
a loop over all intra-cell interactions is used. But instead of using a static loop to start the inter-cell interaction tasks recursion is used. After completed execution, each intra-cell interaction task calls a inter-cell task with one of its neighbours. This again calls the next inter-cell task, until all neighbours (whole stencil) are evaluated. By this dynamic approach, the DAG will be dynamically built during execution, which reduces the number of dependencies at runtime, allowing more tasks to be executed in parallel.

  \item The last method is buffering the data from every neighbor and using a "reduce" step after the calculation to obtain the final result. Therefore, different tasks no longer write to the same memory and are independent from each other. This is a well known 
solution, but may lead to a dramatic increase in the overall memory consumption of the application. So it is very important to figure out if the application will become memory bound.
\end{enumerate}

\section{Evaluation}
\label{sec:evaluation}
The evaluation of our theoretical prediction was done with a test kernel implementing the feature of the link-cell algorithm from MD as described in sections~\ref{sec:straight-forward-implementation} and \ref{sec:improved_strategies}. The measurements were done on the HLRS systems Laki and Hermit using SMPSs 2.4 and the current Ompss version\footnote{The used versions from the git developer repository were Nanos 5eaf9dc11133365124927ee599fc84b283df8392 and Mercurium d41a4f63ae76881fa2120fdd8617a23a7fc6689c.}. While Laki is a NEC cluster with 8~core dual socket Xeon~X5560 nodes, Hermit is a Cray~XE6 system with 32~core dual socket Opteron~6276 nodes.

\begin{figure}
  \centering
  \includegraphics[width=\columnwidth]{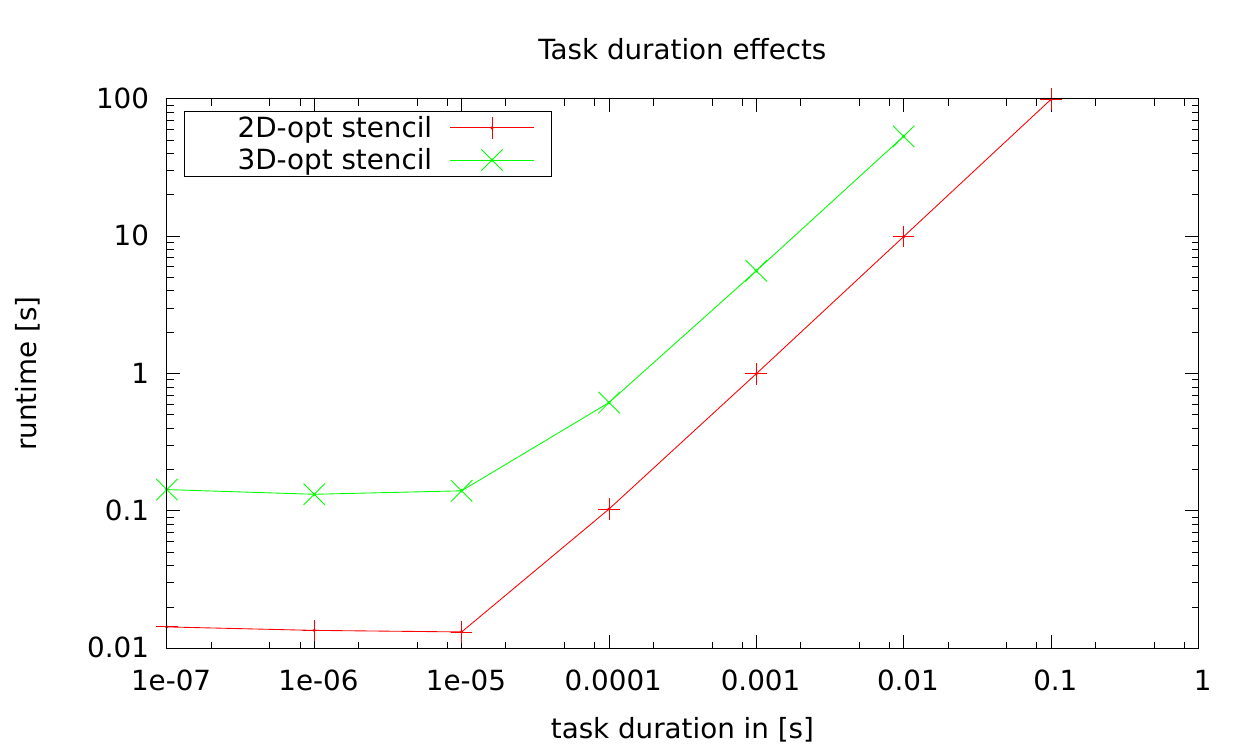}
  \caption{Effect of the single task duration on the overall program runtime using Ompss.}
  \label{fig:taskduration-effects}
\end{figure}

Before the actual evaluation, we measured the influence of the runtime scheduling system overhead. This is important, as we are only interested in the limitation caused by the structure of the task DAG. For this, we varied the
execution time of the single tasks from $100\text{ns}$ to $100\text{ms}$ (see Fig.~\ref{fig:taskduration-effects}).
Based on the results, a single task durations of $5\text{ms}$ was selected for the 3D and $10\text{ms}$ for the 2D examples. These lie well within the region where the runtime overhead can be neglected.

To prevent influences from boundary effects resulting in circular dependencies within the task DAG a $50 \times 50$ cell domain was used for the 2D and a $30 \times 10 \times 10$ cell domain for the 3D experiments.

\begin{figure}[h]
  \centering
  \includegraphics[width=\columnwidth]{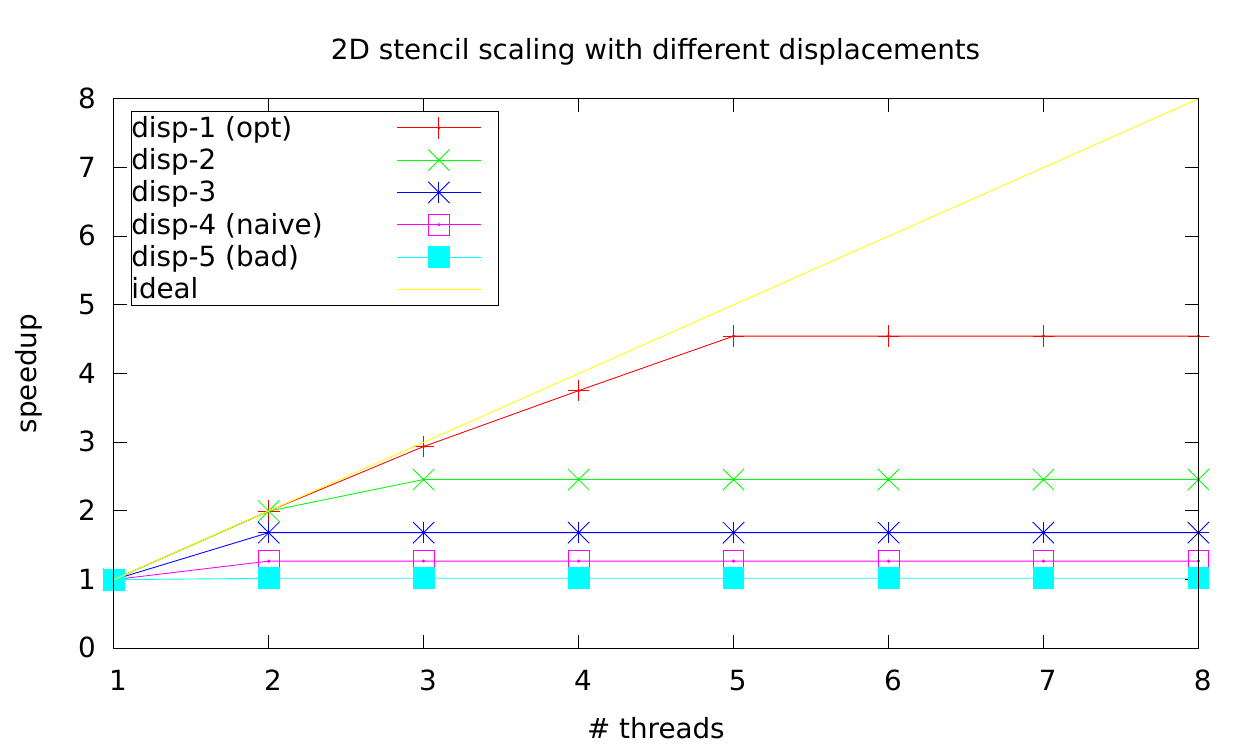}
  \caption{Scaling for different displacements in the 2D stencil obtained with the SMPSs-2.4 runtime on a single HLRS Laki node.}
  \label{fig:2D-scaling-displacement}
\end{figure}

First we implemented the basic algorithm of the 2D and 3D stencil with different interaction orderings and by this resulting in different displacements $\Delta$. As the results in Figure~\ref{fig:2D-scaling-displacement} show, the theoretical values from equation~\eqref{eq:speedup-stencil} are in excellent agreement with the  experimental values.

\begin{figure}[h]
  \centering
  \includegraphics[width=\columnwidth]{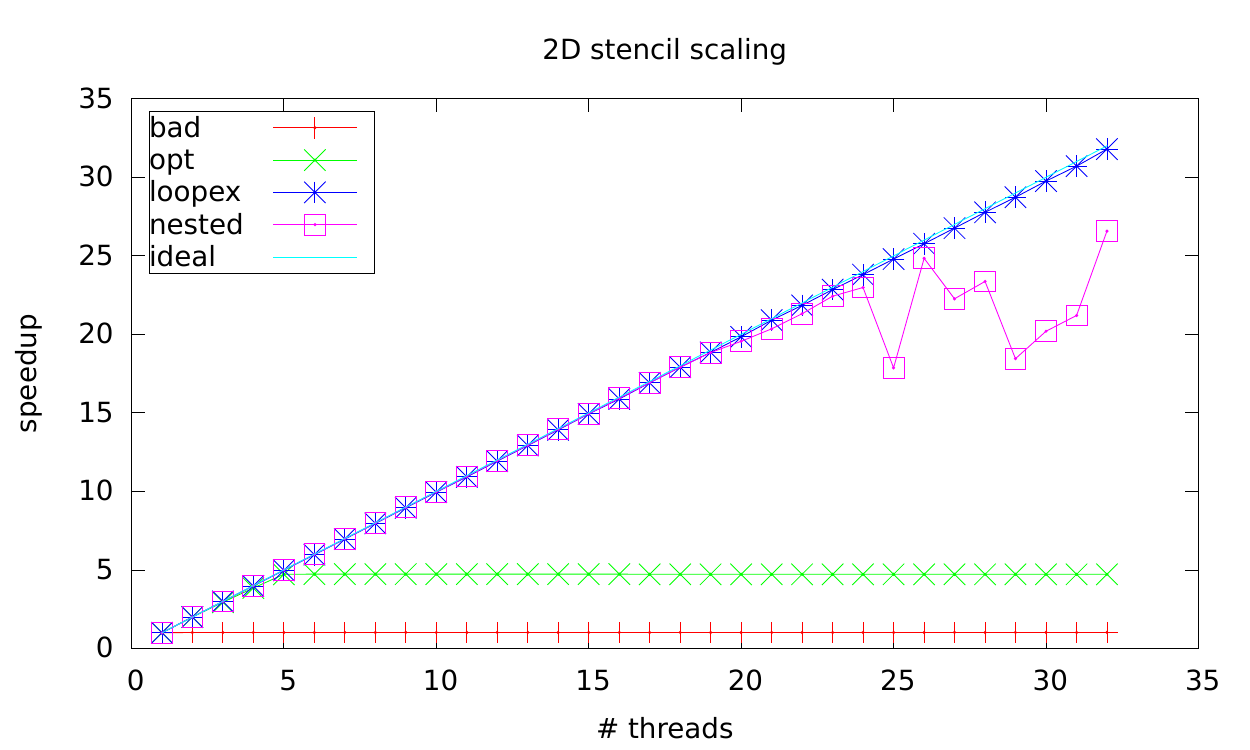} 
  \includegraphics[width=\columnwidth]{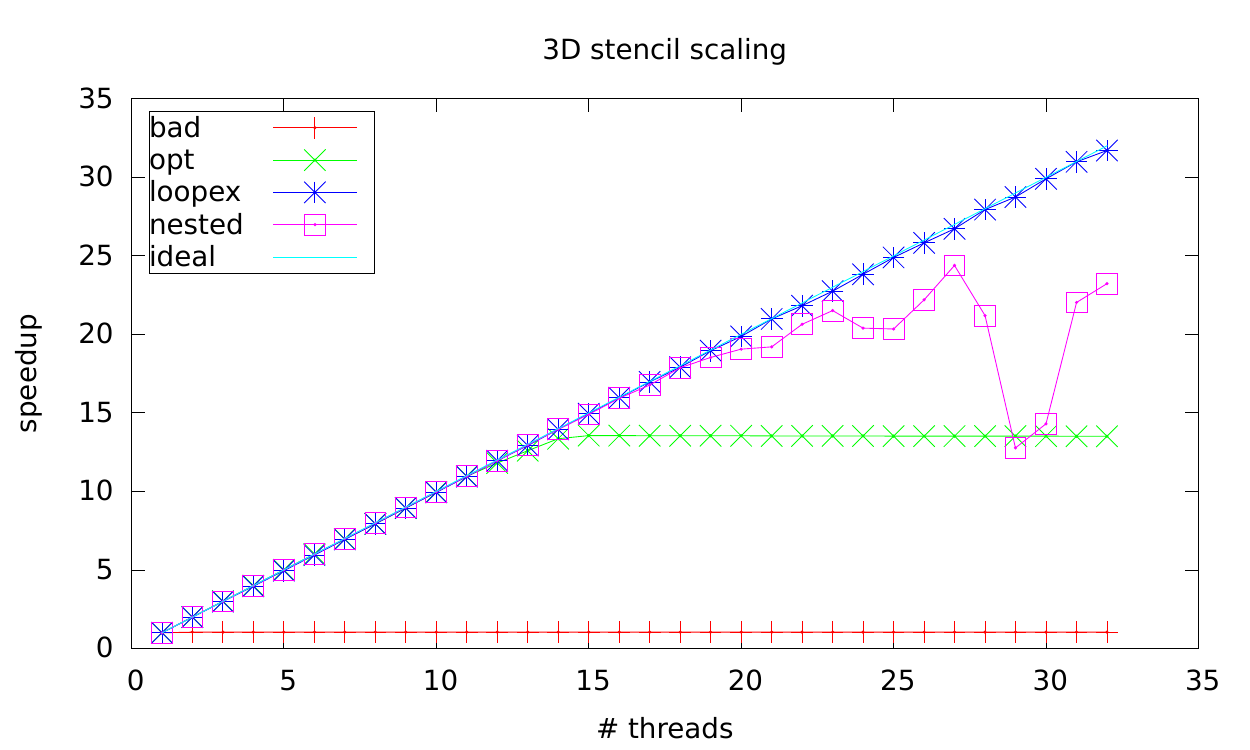} 
  \caption{Scaling of four different kernels: serializing stencil (bad), optimal stencil (opt), loop reordered (loopex) and nested for 2D and 3D on a single Hermit node.}
  \label{fig:optimized-stencil-scaling}
\end{figure}

To evaluate our optimization strategies we implemented the loop exchange and nested versions for the 2D and 3D stencil. The results are shown in Figure~\ref{fig:optimized-stencil-scaling}.

Figure~\ref{fig:optimized-stencil-scaling} shows again perfect agreement with the predicted speedups for the stencil implementations (bad,opt). We see also that the loop exchanged version shows perfect scaling up to 32 threads.
Up to 23 threads we see nearly perfect scaling of the nested version. Above the behavior becomes slightly jittery. The cause for this is probably related to the runtime and currently under investigation by the Ompss developers.

\section{Conclusion}
\label{sec:conclusion}

In this paper we discussed serialization effects in data dependency driven task
parallel algorithms for spatial decomposition. As an
example, we showed that the implementation of the link-cell
algorithm described in literature, with concurrent cell updates, results in a limited speedup. In this
 algorithm, the order in which the cells in the neighbour stencil
are evaluated has a direct effect on the speedup.
Starting from the task DAG and the critical path we derived the theoretical
upper limit for the speedup and verified it experimentally. The cause
for this limitation was identified to be the lack of a mutual exclusion or
concurrent execution for tasks which do update the same data but do not depend
on a specific execution order.
In general it can be seen that the stencil implementation has a mayor impact on
parallelism and the programmer should pay attention to this. For a 3D simulation,
the maximal speedup varies between 1 and 14 depending on the order in which the stencil
is evaluated.

To achieve speedups beyond the theoretical limited given by \eqref{eq:speedup-stencil-max}, we implemented two other kernels resulting
in a wider DAG: a loop and displacement based version as well as a version using nested
tasks.
Both versions are not limited in their theoretical speedup as long as the number of
tasks is large enough. The applicability of those two techniques was again
proven by experiments. The results show pretty good scalability. The loop exchanged + coloring version scales perfectly up to 32~threads. The nested version scales up to 23~threads but shows weird behavior with more threads, which is probably caused by the runtime and under investigation.

As we have shown, dependency driven parallelization of codes based on spatial decomposition with data sharing, 
may easily result in severe and unnecessary speedup limitations. In the worst case, the attempted parallelization 
results in complete serialization.
This can easily be avoided by adhering to simple
rules, as put forth in this paper.
Taking this advice into account the StarSs approach for dependency aware task based programming delivers good performance with little effort for the programmer. After discussing the results presented in this paper with the StarSs developers, the integration of a mutual exclusion clause is now planned.

\section*{Acknowledgments}
This work was supported by the European Community's Seventh
Framework Programme [FP7-INFRASTRUCTURES-2010-2] project TEXT under
grant agreement number 261580 and project APOS under 277481.

\bibliographystyle{IEEEtran}
\bibliography{IEEEabrv,literature.bib}

\end{document}